\documentclass[sigconf,natbib=true]{acmart}

\AtBeginDocument{%
  \providecommand\BibTeX{{%
    \normalfont B\kern-0.5em{\scshape i\kern-0.25em b}\kern-0.8em\TeX}}}
    
\input{zhwinfonts}
\copyrightyear{2023}
\acmYear{2023}
\setcopyright{acmlicensed}
\acmConference[CIKM '23] {Proceedings of the 32nd ACM International Conference on Information and Knowledge Management}{October 21--25, 2023}{Birmingham, United Kingdom.}
\acmBooktitle{Proceedings of the 32nd ACM International Conference on Information and Knowledge Management (CIKM '23), October 21--25, 2023, Birmingham, United Kingdom}
\acmPrice{15.00}
\acmISBN{979-8-4007-0124-5/23/10}
\acmDOI{10.1145/3583780.3615210}
\usepackage[utf8]{inputenc}
\usepackage{amsmath}
\usepackage{tablefootnote}

\usepackage[utf8]{inputenc}
\usepackage{graphicx}
\usepackage{amsmath}
\usepackage{algorithm}
\usepackage{algorithmic}
\usepackage{CJKutf8}
\usepackage{subcaption}
\usepackage{balance}
\usepackage{multirow}
\usepackage[skip=2.5pt]{caption}
\usepackage{array}
\usepackage{booktabs}
\usepackage{threeparttable}
\usepackage{soul}
\usepackage{makecell}

\newcommand{\Z}{\mathcal{Z}}

\newcommand{\eg}{\textit{e.g.}}
\newcommand{\ie}{\textit{i.e.}}
\newcommand{\etal}{\textit{et al. }}

\newcommand\blfootnote[1]{%
  \begingroup
  \renewcommand\thefootnote{}\footnote{#1}%
  \addtocounter{footnote}{-1}%
  \endgroup
}

\begin{document}



\title[Differentiable Retrieval Augmentation via Generative Language Modeling]{Differentiable Retrieval Augmentation via Generative Language Modeling for E-commerce Query Intent Classification}

\author{Chenyu Zhao$^\dag$}
\email {zhaochenyu8@jd.com}
\affiliation{%
  \institution{\small JD.com}
  \country{}
}
\author{Yunjiang Jiang$^\dag$}
\email {yunjiangster@gmail.com}
\affiliation{%
  \institution{\small JD.com}
  \country{}
}
\author{Yiming Qiu}
\email {qiuyiming3@jd.com}
\affiliation{%
  \institution{\small JD.com}
  \country{}
}
\author{Han Zhang}
\email {zhanghan33@jd.com}
\affiliation{%
  \institution{\small JD.com}
  \country{}
}
\author{Wen-Yun Yang$\,^{*}$}
\email {wenyun.yang@gmail.com}
\affiliation{%
  \institution{\small JD.com}
  \country{}
}

\renewcommand{\shortauthors}{Chenyu Zhao, Yunjiang Jiang, Yiming Qiu, Han Zhang, \& Wen-Yun Yang}


\begin{abstract}
Retrieval augmentation, which enhances downstream models by a knowledge retriever and an external corpus instead of by merely increasing the number of model parameters, has been successfully applied to many natural language processing (NLP) tasks such as text classification, question answering and so on. However, existing methods that separately or asynchronously train the retriever and downstream model mainly due to the non-differentiability between the two parts, usually lead to degraded performance compared to end-to-end joint training.
In this paper, we propose \textbf{D}ifferentiable \textbf{R}etrieval \textbf{A}ugmentation via \textbf{G}enerative l\textbf{AN}guage modeling (\textit{Dragan}), to address this problem by a novel differentiable reformulation.
We demonstrate the effectiveness of our proposed method on a challenging NLP task in e-commerce search, namely query intent classification. 
Both the experimental results and ablation study show that the proposed method significantly and reasonably improves the state-of-the-art baselines on both offline evaluation and online A/B test.

\blfootnote{$^\dagger\,$ Both authors contribute equally}
\blfootnote{$^*\,$ corresponding author}

\end{abstract}


\maketitle

\section{Introduction}

In the recent few years, large natural language processing (NLP) models~\cite{devlin2018bert, raffel2020exploring, lin2021m6, chowdhery2022palm, radford2018improving} have emerged as a breakthrough approach to many long-standing NLP problems, such as question answering~\cite{yang2019end, wang2019multi}, text classification~\cite{sun2019fine, lu2020vgcn}, entity extraction~\cite{chang2021chinese, 2015Biomedical}, semantic retrieval~\cite{huang2013learning, zhang2020towards, zhang2021joint} and so on. The number of parameters in those large NLP models have exploded from tens of millions~\cite{devlin2018bert}, hundreds of millions~\cite{raffel2020exploring}, to a stunning hundreds of billions~\cite{openai2023gpt4}. 
However, it is still worth re-thinking if the formidable scale of the models is the final destiny of NLP, or broadly artificial intelligence (AI)~\cite{openai_large_model_era_gone}. At the same time, some researchers have started another direction of research, namely retrieval augmentation, which essentially utilizes external plugin modules, \eg, a retrieval module, to increase model capacity, instead of merely increasing the number of model parameters.
As two of most representative works, ORQA~\cite{lee2019latent} proposes a static knowledge retriever that can retrieve from external corpus to help question-answering task, and REALM~\cite{guu2020realm} enhances downstream task by language model pre-training with a knowledge retriever.
These retrieval augmentation approaches provide more appealing solutions to existing NLP problems, since they allow external corpus to be used as plugin module, which is much more scalable and flexible than model parameters.

In this paper, we explore retrieval augmentation for a challenging problem in e-commerce search, namely query intent classification.
In a typical e-commerce system, both items and queries are categorized into multi-level hierarchical structure of categories~\cite{qiu2022pre} in order to be utilized later in multiple scenarios, such as product search, targeted marketing, inventory management, sales analysis and so on. 
Despite playing a vital role in search pipeline, query intent classification remains one of the most challenging tasks in e-commerce search, since user queries are usually short, polysemous, and lack of training labels especially for long tail queries. 
As shown in Table~\ref{tab:hard_example}, we can see a few query examples that are nearly impossible to classify correctly without leveraging external data sources: 1) query ``lomography'' is quite a niche brand in photography, which occurs only a handful of times in the whole websites and zero times in our training data.
2) Query ``BCD-253WDPDU1'' and ``huawei HD98SOKA'' correspond to specific refrigerator and television models, respectively. 
These queries are literally meaningless, but they make sense if we can retrieve some items with matched tokens to ``augment '' the query.

\begin{table*}[ht]
    \caption{Examples of hard queries that need retrieval augmentation.}
    \vspace{-1mm}
    \label{tab:hard_example}
    \centering
    \resizebox{\textwidth}{!}{
    \begin{tabular}{c|c|c}
    \hline
    Query & Target Item & Category \\
    \hline
    lomography & LOMOGRAPHY Lomo’Instant Automat Polaroid Camera Retro Red Leather Limited Edition with three lenses & Photography \\
    BCD-253WDPDU1 & Haier/Haier refrigerator three-door 253-liter frequency conversion air-cooled frost-free household refrigerator BCD-253WDPDU1 & Refrigerator \\
    huawei HD98SOKA & HUAWEI Smart Screen V98 98-inch HD98SOKA 120Hz anti-glare giant screen 4K ultra-high-definition eye-protection smart gaming TV & Television \\
    \hline
    \end{tabular}}
    \vspace{-3mm}
\end{table*}

However, existing retrieval augmentation methods, though with initial successes, still suffer from a few deficiencies: 
1) the retrieval index update is non-differentiable, which prevents the model to be trained in end-to-end manner. 2) The model training is too expensive, mostly due to the non-differentiable index building. Thus the index has to be re-built many times during the retrieval model training. 3) 
Retrieving the whole document seems unnecessary and it introduces more computation burden for downstream models. Thus, as an analogy to human memory, we are very interested in exploring if the retrieval module can just generate some useful ``fragments'' instead of the whole document, since we believe that's how our human thinks.

To address the above deficiencies, in this paper we propose a differentiable retrieval augmentation model to jointly learn a neural knowledge retriever and a knowledge-augmented classifier for query intent classification.

\section{Related Work}
\subsection{Query Intent Classification}

Over the last decade, deep learning approaches have been widely applied in query intent classification due to their capability of semantic representation~\cite{hashemi2016query,chen2019bert}, for example, transfer learning~\cite{howard2018universal,skinner2019commerce}, external knowledge utilization (such as Wikipedia~\cite{hu2009understanding}, search content~\cite{ashkan2009classifying} and knowledge base~\cite{wang2015query}), and retrieval augmentation at a small scale~\cite{broder2007robust}. Recently, BERT~\cite{devlin2018bert} based methods significantly improve a lot of NLP tasks, including query classification~\cite{chen2019bert} and document classification~\cite{cai2021slim}.


\subsection{Retrieval Augmentation}
Weston \etal~\cite{cai2019retrieval} first introduces retrieval augmentation technique into dialogue response generation domain, and Li \etal~\cite{li2022survey} provides a survey about applying retrieval augmentation for text generation from some retrieved references instead of generating from scratch.
Recent advances in language model pre-training have shown to be significantly effective in most NLP tasks, including BERT~\cite{devlin2018bert}, T5~\cite{raffel2019exploring} and so on, where the knowledge is encoded implicitly as parameters of the neural network. On the other hand, REALM~\cite{guu2020realm} learns a parameterized retriever to facilitate the main model where the knowledge is encoded explicitly in the external corpus. RECO~\cite{iscen2023retrieval} enhances vision-text models by incorporating a retrieval model to refine their embeddings with cross-modal retrieved information.

\section{Method}
We formulate the query intent classification problem as two sequential tasks: a \textit{neural knowledge retriever} and a \textit{knowledge-augmented classifier}. The former task aims to retrieve supplementary knowledge for the query and the latter task achieves the goal of query intent classification based on the augmented knowledge and input query. In this section, we begin by introducing the details of these two tasks, followed by an empirical and novel training technique, \ie, fragment generation. Finally, we provide a comprehensive overview of the training stages and explain the joint training for the proposed Dragan model.
\begin{figure*}[!ht]
    \centering
    \includegraphics[width=0.8\textwidth]{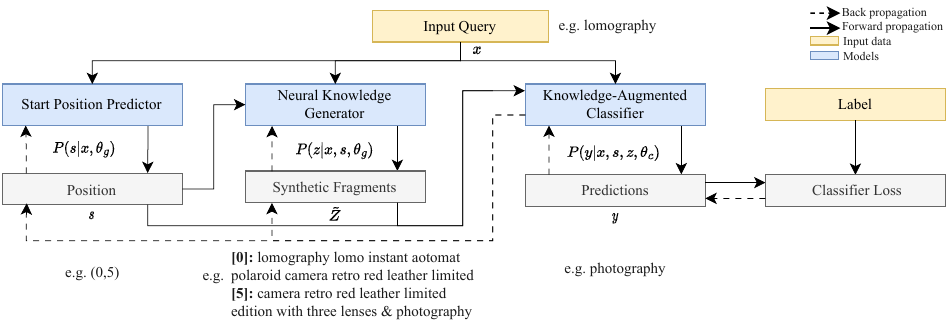}
    \caption{Illustration of the proposed method where three networks work together to produce the final classification result. }
    \label{fig:flow_diagram}
    \vspace{-3mm}
\end{figure*}

\subsection{Query Intent Classification}
\label{sec:query_intent}
For a standard classification problem, we take some input $x$ and learn a distribution $p(y|x)$ over all possible class label $y$. With retrieval augmentation, we have two key components: the \emph{neural knowledge retriever}, which models the generative probability $p(z|x)$, for some intermediate text $z$, and the \emph{knowledge-augmented classifier}, which models $p(y| z, x)$. Thus, the overall log likelihood of $y$ given $x$ can be defined as follows 
\begin{equation}
    \log p(y|x) = \log \sum_{z\in \Z}p (y|z, x) p (z | x).
    \label{eq:likelihood}
\end{equation}


\textbf{Neural Knowledge Retriever.} 
Previously, REALM~\cite{guu2020realm} propose knowledge retrieval as an embedding retrieval problem with inner product as the similarity measure. The results are promising but the indexing part is cumbersome -- the embedding index has to be asynchronously updated during the training, because the retrieval is based on inner product and there is no easier way to find $k$ nearest neighbors other than rebuilding the embedding index. 

Here we explore another approach to knowledge augmentation based on text generation by replacing the retrieval probability $p(z | x)$ with text generative probability, which consequentially gets rid of the embedding index and its cumbersome updating. Formally, the differentiable retriever can be defined as
\begin{equation}
    p(z|x; \theta_g) = \Pi_{w_i \in z, 1\leq i \leq n-1}p(w_{i+1}|w_{i}; \theta_g).
\end{equation}
where $n$ is the length of text $z$, $p(w_{i+1}|w_{i})$ is learned by a transformer-based model, and $p(z|x)$ is the text generative probability instead of retrieval probability, which can be learned jointly and synchronously with the following classification step. However, one may notice that it incurs one another challenge -- the text generator usually involves $\arg\max$ operator when decoding tokens, which is non-differentiable. Next we present a technique to circumvent it.


\textbf{Knowledge-augmented Classifier.} 
With the input $x$ and the generated text $z$, we can get the label prediction by the probability $p (y|z, x)$.
Since it is prohibitively expensive to compute the summation in Equation~(\ref{eq:likelihood}) due to the exponential search space of $\Z$, instead we propose to use a subset of the full space $\tilde{\Z} =\{{z_1,z_2,\cdots,z_k}\}   \subset \Z$ to approximate it.
\begin{equation}
    \log p(y|x) \approx \log \tilde{p}(y | x) := \log \sum_{z\in \tilde{\Z}}p (y|z, x; \theta_c) p (z | x; \theta_g).
    \label{eq:approx}
\end{equation}
In practice, we use the top-$k$ set of documents $z_i$ by the generative model for the given query $x$. As $k \ll |\Z|$, it is feasible to calculate the gradient for Equation~(\ref{eq:approx}) efficiently by enumerating all $z_i$ in $\tilde{\Z}$ as follows
\begin{eqnarray*}
    \cfrac{\partial \log \tilde{p}(y|x)}{\partial \theta_c} &=&  \cfrac{\sum_{z\in \tilde{\Z}} \cfrac{\partial p (y|z, x; \theta_c)}{\partial \theta_c} p (z | x; \theta_g)}{\sum_{z\in \tilde{\Z}}p (y|z, x; \theta_c) p (z | x; \theta_g)},\\
    \cfrac{\partial \log \tilde{p}(y|x)}{\partial \theta_g} &=&  \cfrac{\sum_{z\in \tilde{\Z}} p (y|z, x; \theta_c) \cfrac{\partial p (z | x; \theta_g)}{\partial \theta_g}}{\sum_{z\in \tilde{\Z}}p (y|z, x; \theta_c) p (z | x; \theta_g)}.
\end{eqnarray*}
While these approximate gradients may differ significantly from the true gradients $\nabla \log p(y |x)$, Eq.~\eqref{eq:approx} implies that optimizing $\tilde{p}(y|x)$ approximately optimizes $p(y|x)$ as well.
Note also that pre-training the generator $\theta_g$ ensures that the approximation in Eq.~\eqref{eq:approx} holds from the beginning. Pre-training the classifier $\theta_c$ is less necessary, though in practice it also gives small improvement.

\subsection{Fragment Generation}
\label{sec:fragment}
In practice, straightforward using generative model as retrieval augmentation can be highly inefficient: 1) the sequential decoding process during text generation cannot be parallelized~\cite{graves2013generating}.
2) With full length of e-commerce item title, which is normally around 100 characters~\cite{qiu2021query}, the final classification model also suffers from the $O(n^2)$ complexity of the self attention layer in the transformer models~\cite{vaswani2017attention}. Besides efficiency, longer text generation can also lead to semantic drift and degeneration~\cite{holtzman2019curious, merity2017regularizing}, affecting downstream classification accuracy.
Thus, we come up with the following fragment generation method to replace the full item title generation.

Formally, we first choose a set $\tilde{\mathcal{S}}$ of fragment start positions $s$ according to input text $x$ by a transformer encoder, which shares same parameters as the generator encoder. Instead of directly obtaining the top-$k$ start positions, we sample some start positions based on their probabilities. This approach helps us explore long-tail start positions that may be useful for classification. Gumbel-softmax~\cite{jang2016categorical} is applied to allow pass-through back-propagation. 
\begin{equation*}
    p(s|x) = \rm{Softmax}(\rm{Encoder}(x)) -\log(-\log(\boldsymbol{\textit{U}}(0, 1))).
    \label{eq:startpos}
\end{equation*}

Next, we generate a fixed number of fragment tokens step by step starting from position $s$. The generation probability $p(z|x)$ in Equation (\ref{eq:likelihood}) can be rewritten as
\begin{equation*}
    p(z|x) = \sum_{s\in \tilde{\mathcal{S}}}p(z|x,s)p(s|x),
    \label{eq:frag_probs}
\end{equation*}
and the overall log likelihood can be rewritten as
\begin{equation}
    \log p(y|x) \approx \log \sum_{z\in \tilde{\Z}}p (y|z, x; \theta_c) \sum_{s\in \tilde{\mathcal{S}}} p(z | x, s; \theta_g) p (s | x; \theta_g).
    \label{eq:frag_likelihood}
\end{equation}
The gradient with respect to generator and classifier parameters can be rewritten similarly, without impact on the differentiability.

In practice, we find the fragment generation is good enough for our latency requirements, with better accuracy compared to full text generation (see Experiment, Section~\ref{sec:ablation}).

\subsection{Model Training Details}
Our training process involves two stages: pre-training and joint training. In the first stage, we utilize a dataset of user-clicked log (query, title, category) to pre-train the generation model. This model generates fragments of product titles based on the query and a randomly selected start position, on top of which we pre-train the classification model. In the second stage, we jointly fine-tune the generation model with the classification model. Specifically, we first sample start positions based on the predicted probabilities and employ a beam search decoder~\cite{freitag2017beam} to predict fragment tokens based on the sampled start positions. In our setting, we choose top-5 start positions, and for each start position, we use beam size of 3, thus resulting in a total of 15 augmented fragments for each query. Finally we concatenate the input query with the augmented fragments for classification. Note that the generation model continues to update throughout the joint training stage. Moreover, to prevent over-fitting, we randomly initialize the last dense layer of pre-trained classification model.


Our vocabulary consists of around 9,000 frequently used Chinese characters and word pieces from the English alphabet, along with other special tokens. We do not use the conventional BERT vocabulary of around 20,000 tokens, since smaller vocabulary significantly improves the decoder efficiency and GPU memory usage.



\section{Experiment}

\subsection{Setup}
\subsubsection{Dataset}
As shown in Table~\ref{tab:dataset}, we use a publicly available dataset~\cite{qiu2022pre} for the sake of reproducibility, which is collected from user click log in a major e-commerce search platform.


\begin{CJK*}{UTF8}{gbsn}
\begin{table*}[htb]
    \centering
    \caption{Start positions and generated title fragments of different queries.}
    \resizebox{0.98\textwidth}{!}{
    \begin{tabular}{c|c|c|c}
        \hline
        Query & Related item title & Start position & Generated fragment \\
        \hline
        \multirow{3}{*}{\makecell[c]{27英寸2k \\ (27-inch 2k)}} & \multirow{3}{*}{\makecell[c]{HKC 27英寸 2K高清144Hz电竞 1800R曲面屏幕 hdmi吃鸡游戏 不闪屏 支持壁挂 液晶电脑显示器 SG27QC \\ (HKC 27-inch 2K High Definition 144Hz Gaming Monitor with 1800R Curved Screen, HDMI for Chicken Dinner Games, \\ Non-Flickering, Wall Mountable, LCD Computer Display SG27QC)}} & 51 & 显示器Q2790P (Monitor Q2790P) \\
         & & 45 & 电脑显示器显示屏 (Computer monitor) \\
         & & 49 & 电脑显示器SG27 (Computer monitor SG27) \\
        \hline
        \multirow{3}{*}{b550} & \multirow{3}{*}{\makecell[c]{玩家国度（ROG）ROG STRIX B550-A GAMING吹雪主板支持 CPU 5600G/5600X/5700G \\ (ROG STRIX B550-A GAMING Blizzard motherboard supports CPU 5600G/5600X/5700G)}} & 25 & IFI迫击炮电脑主板 (IFI Mortar computer motherboard) \\
         & & 28 & MING吹雪主板支持 (MING Blizzard motherboard support) \\
         & & 26 & GAMING吹雪主板 (GAMING Blizzard motherboard)  \\
        \hline
    \end{tabular}}
    \label{tab:case}
    \vspace{-3mm}
\end{table*}
\end{CJK*}

\begin{table}[t]
    \centering
    \caption{Dataset statistics.}
    \resizebox{0.95\columnwidth}{!}{
    \begin{tabular}{c|cccc}
        \hline
        Dataset   & \# Examples &  \# Queries & \# Items &  \# Categories  \\ \hline
        Pre-training   & 931,429   & 200,001 & 101,207 & 98  \\ 
        Joint-training   & 667,665 & 180,008 & 83,672  & 96 \\ 
        Overall eval  & 10,000   & 10,000 & --- & 94    \\ 
        Long-tail eval  & 10,000   & 10,000 & --- & 88 \\
        \hline
    \end{tabular}}
    \label{tab:dataset}
\end{table}

\subsubsection{Evaluation Metrics}
We use precision (P), recall (R), f1 score (F1) as our evaluation metrics, where the precision measures the accuracy of predicted query categories, the recall measures the proportion of correctly predicted categories.

\subsubsection{Baseline Methods}
We compare our proposed method Dragan with the below four baselines:
\begin{itemize}
    \item \textbf{ORQA}~\cite{lee2019latent} stands for the model released by Google research, which contains an inverse cloze task(ICT) embedding retrieval model and a downstream model. In our setting, the downstream model is a BERT classifier. Note that the retrieval model does not update in the joint training stage.
    \item \textbf{REALM}~\cite{guu2020realm} stands for a pre-training method evolved from ORQA model, where the retrieval task is jointly trained with a token prediction task. Here we also apply the same BERT model for classification, as the proposed Dragan method. 
    \item \textbf{Query-only} stands for the BERT model trained directly on the dataset without any pre-training. 
    \item \textbf{RSC}~\cite{qiu2022pre} stands for the model pre-trained with randomly clipped item titles and fine-tuned on queries.
\end{itemize}

\subsection{Results}


Table~\ref{tab:comparison} shows the comparative results between the proposed Dragan with the above baseline methods, from where we can observe that our 12-layer Dragan model improves all baselines by a large margin in terms of F1 metrics. 
Moreover, since we aim at a more practical scenario where we can serve the model online even with CPU machines, we also include the results of Dragan with a 4-layer generator and a 2-layer classifier. Surprisingly, this much smaller Dragan still significantly improves the other baseline methods, especially on long-tail queries. 

\begin{table}[t]
\centering
\caption{Comparative results with baseline methods.}
\label{tab:comparison}
\footnotesize
\resizebox{0.95\columnwidth}{!}{
\begin{tabular}{c|ccc|ccc}
\hline
\multirow{2}{*}{Method} & \multicolumn{3}{c|}{Overall} & \multicolumn{3}{c}{Long-tail} \\
\cline{2-7}
 & P & R & F1 & P & R & F1 \\
\hline
Query-only (-, 12) & 0.680 & 0.860 & 0.683 & 0.595 & 0.939 & 0.656 \\
ORQA (12,12) & 0.871 & 0.746 & 0.764 & 0.682 & 0.830 & 0.690 \\
REALM (12,12) & 0.874 & 0.750 & 0.767 & 0.625 & 0.840 & 0.638 \\
RSC (-,12) & 0.818 & $\mathbf{0.864}$ & 0.787 & 0.737 & $\mathbf{0.951}$ & 0.782 \\
\hline
Dragan (4,2) & 0.867 & 0.813 & 0.800 & 0.795 & 0.911 & 0.820 \\
Dragan(12,12) & $\mathbf{0.896}$ & 0.811 & $\mathbf{0.816}$ & $\mathbf{0.856}$ & 0.920 & $\mathbf{0.868}$ \\
\hline
\end{tabular}}

\begin{tablenotes}
\item[*] *$\,$The (x, y) notation stands for x-layer retriever and y-layer classifier.
\end{tablenotes}
\vspace{-2mm}
\end{table}

\subsection{Ablation Studies}
\label{sec:ablation}
In Table~\ref{tab:ablation}, we compare the different variants of the proposed model: 1) \emph{full title} stands for generating the full item title instead of just fragments introduced in Section~\ref{sec:query_intent}, 2) \emph{equal probability} stands for using equal probabilities for all fragments, instead of using learnable generation probabilities, 3) \emph{fixed generator} stands for freezing the generator parameters while training the classifier, which means no joint-training.


We can make a few observations from the comparative results: 1) using fragment generation improves full title generation  by 2.2\% and 2.7\% in F1 metric for overall and long-tail evaluation, which may be caused by the noisy tokens in the full item title. Additionally, the inference time of fragment generation is much shorter than the full title generation, which enables us to serve it online. 2) Using learned probability improves the equal probability by 2.3\% and 2.5\% in F1 metric for overall and long-tail evaluation
. 3) Joint training of the generator and classifier significantly improves the fixed generator, especially on long-tail evaluation. This demonstrates the necessity of joint training, enabled by our proposed differentiable retriever.

\begin{table}[t]
\centering
\caption{Comparison between model variants.}
\label{tab:ablation}
\small
\setlength{\tabcolsep}{1.7mm}
\resizebox{0.95\columnwidth}{!}{
\begin{tabular}{c|ccc|ccc|c}
\hline
\multirow{2}{*}{Ablation} & \multicolumn{3}{c|}{Overall} & \multicolumn{3}{c|}{Long-tail} & \multirow{2}{*}{\makecell[c]{Inference \\ time(ms)}} \\
\cline{2-7}
 & P & R & F1 & P & R & F1 & \\
\hline
Dragan (4,2) & 0.867 & 0.813 & 0.800 & 0.795 & 0.911 & 0.820 & 149 \\
Full title & 0.833 & 0.809 & 0.778 & 0.759 & 0.909 & 0.793 & 569 \\
Equal probability & 0.838 & 0.800 & 0.777 & 0.763 & 0.902 & 0.795 & -- \\
Fixed generator & 0.846 & 0.815 & 0.786 & 0.772 & 0.916 & 0.804 & -- \\
\hline
\end{tabular}}
\end{table}

\begin{figure}[t]
    \vspace{-4mm}
    \centering
    \includegraphics[width=0.40\textwidth]{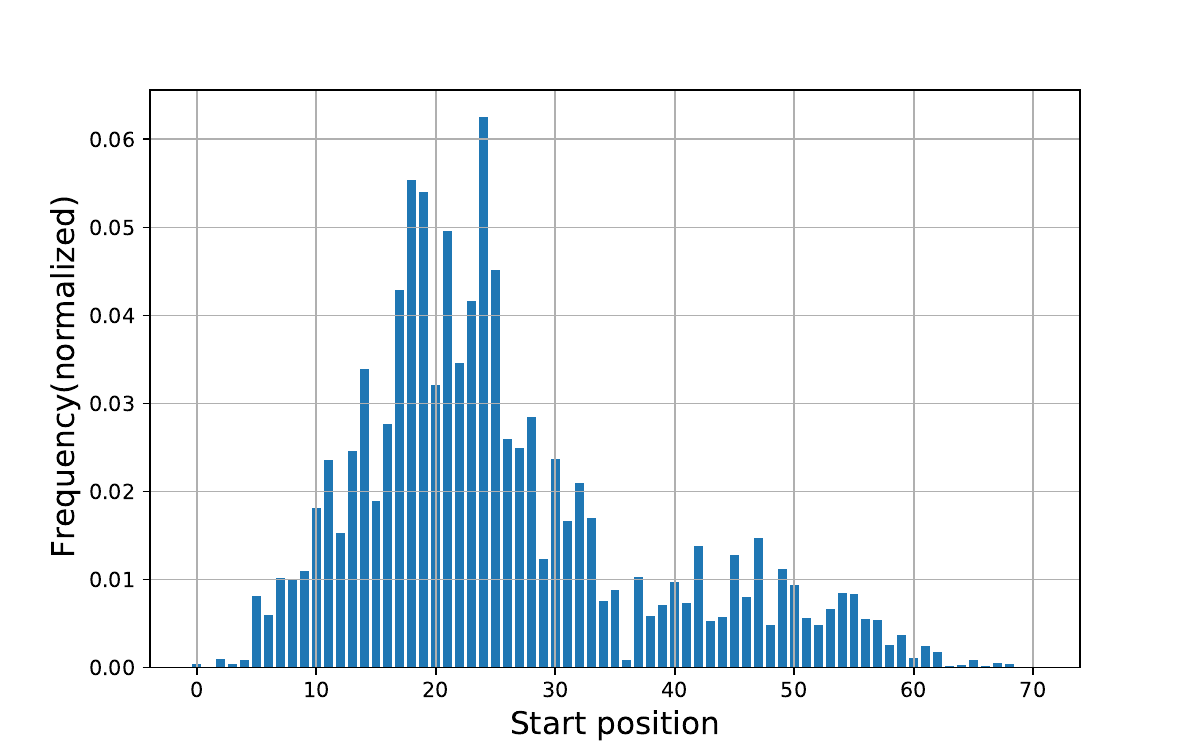}
    \caption{Start position distribution on overall evaluation.}
    \label{fig:start_position}
    \vspace{-4mm}
    
\end{figure}


In Figure~\ref{fig:start_position}, we present the start position distribution of the overall evaluation set, which shows that most of the start positions are concentrated in the range of 10-30. 
Table~\ref{tab:case} shows that the model tends to predict the positions where the product words are located, thus most generated fragments contain the product words. This is highly intuitive as it aligns with how humans understand a query. 


\subsection{Online Performance}
We conduct A/B test on a leading e-commerce search system, using 20\% of the entire site traffic during a period of 30 days. The online query intent classification baseline is the RSC~\cite{qiu2022pre} model.
Due to the confidential business information protection, we only report the relative improvements, where the gross merchandise value (GMV), the number of unique order items per user (UCVR), and the click-through rate (CTR) are significantly improved by 0.13\%, 0.28\% and 0.08\%, respectively.

\section{Conclusion}
We have introduced a differentiable generator-based retrieval augmentation approach that can be jointly trained with any downstream tasks, which fixes a major weakness in existing retrieval augmentation methods. We apply the proposed method to an e-commerce query intent classification problem, and achieve significant improvements over previous methods for both offline and online evaluations. At last, we want to emphasize that the proposed approach is general enough to be applied to other NLP tasks.

\clearpage
\bibliographystyle{ACM-Reference-Format}
\balance
\bibliography{references}

\end{document}